\documentclass[aps,prl,twocolumn,showpacs]{revtex4}
\usepackage{epsfig}
\newcommand{\be}{\begin{equation}} 
\newcommand{\ee}{\end{equation}}
\newcommand{\bea}{\begin{eqnarray}} 
\newcommand{\eea}{\end{eqnarray}}

\newcommand{\gton}{\mathrel{\lower.9ex \hbox{$\stackrel{\displaystyle 
>}{\sim}$}}} 
\newcommand{\lton}{\mathrel{\lower.9ex \hbox{$\stackrel{\displaystyle 
<}{\sim}$}}}

\newcommand{\vp}{{\vec p}}

\newcommand{\vq}{{\vec q}}

\begin{document}
\title{Novel mechanism of high-$p_T$ production 
from an opaque quark-gluon plasma}
\author{D\'enes Moln\'ar}
\affiliation{Department of Physics, Ohio State University, 174 West 18th Ave,
Columbus, Ohio 43210, USA}

\date{\today}
\begin{abstract}
We find that in an opaque quark-gluon plasma,
significant particle production at high transverse momenta 
can occur via acceleration of (initially) lower-momentum partons.
This new mechanism, which is the opposite of parton energy loss 
(jet quenching), has important implications for observables, such as elliptic
flow, in heavy-ion collisions at RHIC energies and above.
\end{abstract}

\maketitle

{\em Introduction. - }
There are several indications that 
an opaque, strongly-interacting quark-gluon plasma has been 
created~\cite{MiklosQGP,EdwardQGP,HorstQGP} in relativistic heavy-ion 
collisions at 
the Relativistic Heavy Ion Collider (RHIC).
One key observable is the large azimuthal momentum anisotropy
of particle production~\cite{STARv2charged,PHENIXv2charged,PHOBOSv2charged}
in noncentral collisions,
characterized by the elliptic flow coefficient 
$v_2\equiv \langle \cos 2\phi \rangle$, 
which requires
more than an order of magnitude higher opacities than that of 
a weakly-coupled perturbative parton plasma~\cite{v2}.
Though such a dense system is still dissipative~\cite{dissipv2}, i.e., it is
{\em not} an ideal fluid,
transport mean free paths $\lambda\sim 0.1$ fm are as short as
 their quantum mechanical lower
bound~\cite{MiklosPawel85}
and, therefore, the system seems to be the most perfect fluid possible.

The large parton opacities have several consequences.
For example, charm quarks and antiquarks 
are predicted to acquire a large elliptic 
flow~\cite{charmv2},
and space-momentum correlations of energetic particles are expected 
to show a strong surface emission pattern~\cite{v2cvsh}.
In this paper we show that particle production at high-$p_T$ is also
affected because {\em there is a considerable probability that a low-momentum
parton gets accelerated to high transverse momenta via multiple scatterings.}

We emphasize that the effect considered here is different from the parton
energy gain
due to detailed balance in a {\em static} thermal 
system~\cite{Elossgain}.
The new source of high-$p_T$ particles here is the collective {\em dynamics}
of the dense parton medium. These particles
would be present in the final state, even
if there were no energetic partons in the initial state.

{\em Covariant parton transport theory} is the incoherent particle limit of the
underlying nonequilibrium quantum theory.
Its main advantages are that i) it treats both the low-momentum bulk sector and
high-momentum rare particles in a self-consistent framework; 
ii) it is not limited to local
equilibrium; and iii) it describes the break-up of the system self-consistently.

We consider here an inelastic extension~\cite{charmv2}
of the transport equations in
Refs.~\cite{ZPCv2,v2,ZPC,nonequil},
in which the on-shell parton phase space densities
$\{ f_i(x,\vp)\}$
evolve with elastic $2\to 2$ and {\em inelastic} $2\to 2$ rates as
\begin{widetext}
\bea
p_1^\mu \partial_\mu f_{1,i} &=& S_i(x, \vp_1) +
\frac{1}{16\pi^2}\sum\limits_{jk\ell} 
\int\limits_2\!\!\!\!
\int\limits_3\!\!\!\!
\int\limits_4\!\!
\left(
f_{3,k} f_{4,\ell} \frac{g_i g_j}{g_k g_\ell} - f_{1,i} f_{2,j}
\right)
\left|\bar{\cal M}_{12\to 34}^{ij\to k\ell}\right|^2 
\delta^4(p_1+p_2-p_3-p_4)
 \ .
\label{Eq:Boltzmann_22}
\eea
\end{widetext}
$|\bar{\cal M}|^2$ is the polarization averaged scattering matrix 
element squared,
the integrals are shorthands
for $\int\limits_a \equiv \int d^3 p_a / (2 E_a)$,
$g_i$ is the number of internal degrees of freedom for species $i$,
while $f_{a,i} \equiv f_i(x, \vp_a)$. 
The source functions $\{S_i(x,\vp)\}$ specify the initial conditions.

To study $Au+Au$ collisions at RHIC at $\sqrt{s_{NN}} = 200$ GeV
with impact parameter $b=8$ fm ($\approx 30\%$ central),
we apply (\ref{Eq:Boltzmann_22}) to a system of massless gluons ($g=16$), 
massless light ($u$,$d$) and strange quarks/antiquarks, 
and charm quarks/antiquarks ($g=6$) with mass $M_c = 1.2$ GeV.
The initial conditions and matrix elements were the same as in~\cite{charmv2}.
To model the strongly-interacting parton system at RHIC~\cite{v2},
parton densities were increased two-fold and parton cross sections
were increased about three-fold to $dN(b{=}0)/d\eta = 2000$ at midrapidity 
and $\sigma_{gg\to gg} = 10$ mb, 
from their perturbative estimates $dN/d\eta \approx 1000$ 
and $\sigma_{gg\to gg}\approx 3$ mb.
All elastic and inelastic leading-order $2 \to 2$
QCD processes were taken into account
and were assumed to be enhanced by the same factor $10/3$.
The initial momentum distributions were taken from 
leading-order perturbative QCD
(with a $K$-factor of 2, GRV98LO PDFs, and $Q^2{=}p_T^2$, while
$Q^2 = \hat s$ for charm),
and the low-$p_T$ divergence in the light-parton jet cross sections 
regulated via a smooth extrapolation below $p_\perp < 2$ GeV.
The transverse density distribution was proportional
to the binary collision
distribution for two Woods-Saxon distributions,
therefore $dN^{parton}(b{=}8\ {\rm fm})/dy \approx 500$.
Perfect $\eta=y$ correlation was assumed.

The transport solutions were obtained via Molnar's Parton Cascade 
algorithm~\cite{MPC} (MPC),
which employs the parton subdivision technique~\cite{subdivision} 
to maintain Lorentz covariance and causality.

{\em Results. - }
There are three general ways a particle can acquire its {\em final} transverse momentum
during the transport evolution:

\smallskip
\noindent
i) it has no interactions at all, 
in which case the particle comes from the {\em corona}
of the collision region;

\smallskip
\noindent
ii) it {\em loses} energy via interactions and moves to lower $p_T$, which is the jet quenching 
component; or

\smallskip
\noindent
iii) it interacts and {\em gains} energy, i.e., gets pushed to higher $p_T$ by 
the medium.

Figure~1 shows the relative fractions of these three components as a function 
of the final
$p_T$ from the transport calculation
for partons at midrapidity ($|y_f|<1$).
At low final $p_T \sim 1-2$ GeV, partons predominantly arrive via the collective 
``push'' from the medium. 
Contributions from quenched partons are smaller, and the corona is negligible. 
On the other hand,
as $p_T$ increases, the quenched component and the corona become more and more 
significant,
while the accelerated part gradually decreases.
However, even at $p_T \sim 7-8$ GeV, energy loss gives only one half of the 
yield,
while almost {\em one-third} of the partons are accelerated (initially) 
lower-momentum partons. 
\begin{figure}[htpb]
\centerline{\epsfig{file=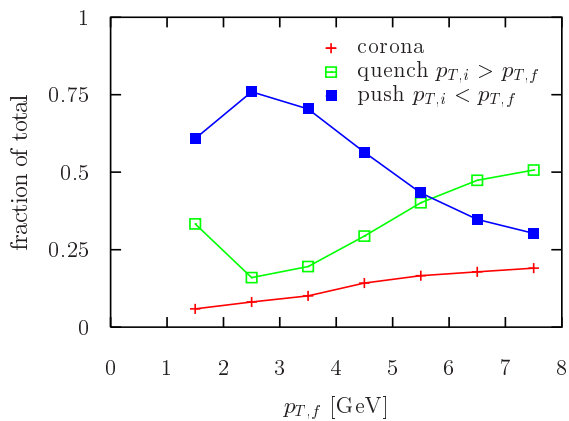,width=2.92in,height=2.2in,angle=0}}
\caption{Fraction of the final high-$p_T$ yield coming from partons without 
interactions (corona),
from energy loss (quench), and from accelerated partons (push)
as a function of $p_T$
in $Au+Au$ at $\sqrt{s_{NN}}=200$ GeV with $b=8$ fm, for $\approx 6$ times higher
opacities than perturbative estimates.
Computed via MPC~\cite{MPC,v2}.
}
\label{fig:1}
\end{figure}

Figure~2 
gives a more detailed picture,
where the normalized initial $p_T$ distributions are shown for partons in 
various final $p_T$ intervals
(and final rapidity $|y_f|<1$),
with contributions from the corona omitted.
For all final $p_T$ intervals,
the accelerated soft partons predominantly come from $p_{T,i} \sim 1$ GeV,
which is precisely where the initial parton $p_T$ distribution (dashed line) has a strong peak.
At the opacities considered here, partons have 
$\langle N_{coll}\rangle \approx 14$ scatterings on average.
Though the likelihood of gaining significant $p_T$, even in so many
scatterings, is small, the initial high yield of low-momentum 
partons multiplied with the low
probability can still compete with the much lower parton yields at high $p_T$.
As the final $p_T$ increases, the relative importance of low-momentum partons 
decreases, in accordance
with the decrease in the fractional ``push'' yield in Fig.~1.
\begin{figure}[htpb]
\centerline{\epsfig{file=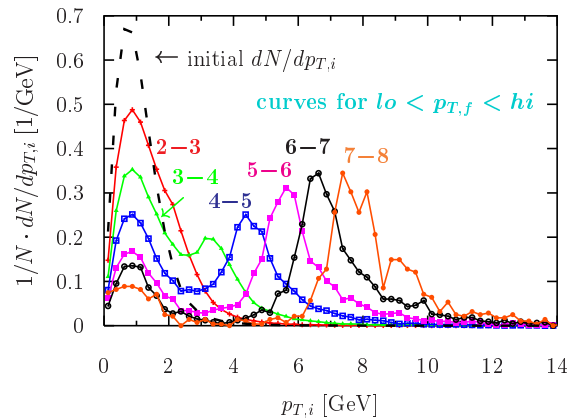,width=2.92in,height=2.2in,angle=0}}
\caption{Normalized initial $p_T$ distributions for partons in various final 
$p_T$ intervals 
(and final $|y_f| < 1$),
omitting contributions from the corona,
from the same calculation as Fig.~1.
The initial parton $p_T$ distribution is also shown (dashed line).
}
\label{fig:2}
\end{figure}

Two-component models based on a thermal (or hydrodynamic) collective
part and a perturbative QCD jet part
underestimate the collective component at high $p_T$ because the 
approximately exponential $p_T$ distribution imposed 
(with typical few-hundred-MeV slope parameters) cannot
compete with the power-law jet spectrum.
However, such a characterization of the collective part is based on macroscopic 
approaches that are formulated in terms of the lowest few moments 
(e.g., particle density, energy density) of the local momentum distribution,
and therefore cannot be trusted for the high-momentum tails. 
Our results in Figs.~1 and 2 imply that (initially) low-momentum partons that 
suffer {\em finite}
number of scatterings populate high transverse momenta with a non-exponential, 
power-law-like distribution.
In fact, already a single scattering generates power-law tails
because for typical Debye-screened $t$-channel gluon exchange
$d\sigma/dt \sim \alpha_s^2/(t-\mu_D^2)^2 \sim \alpha_s^2/(\vq^{\ 2}+\mu_D^2)^2$,
where $\vq$ is the transferred momentum.

The above findings have significant implications for elliptic flow.
Figure~3 shows elliptic flow as a function of $p_T$ for the three components 
(corona, quench, push)
and for their combined yield.
For partons that have lost energy (quench), elliptic flow drops rather sharply 
at high $p_T$,
very similarly to calculations based on inelastic perturbative QCD energy loss 
in the Eikonal limit~\cite{pQCDv2}.
In the transport calculation, however, the sharp drop is 
largely compensated by the much larger $v_2$ of the accelerated partons (push).
The significant ``push'' component also means that the ``geometric'' elliptic 
flow bounds argued in~\cite{shuryaklimit},
which assumed extreme quenching, do not apply.
Though the geometric corona alone gives a positive, $p_T$-independent
elliptic flow 
(reflecting the initial 
spatial anisotropy of the edges of the collision zone), 
it is much smaller than the contribution coming from the 
accelerated partons.
Therefore, the collective dynamics of the opaque quark-gluon plasma
can provide a natural explanation for 
why elliptic flow at RHIC exceeds the ``geometric''
bounds and why it decreases relatively slowly at high transverse 
momenta~\cite{Aihongv2}.
\begin{figure}[htpb]
\centerline{\epsfig{file=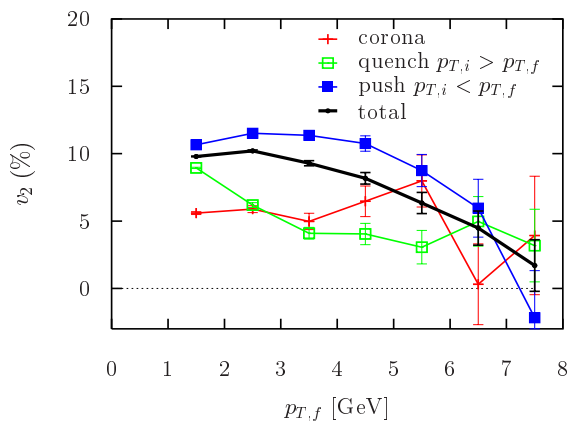,width=2.92in,height=2.2in,angle=0}}
\caption{Elliptic flow as a function of $p_T$ for partons without interactions 
(corona),
from energy loss (quench), from accelerated partons (push),
and from all three components combined (total),
from the same calculation as Fig.~1.
}
\label{fig:3}
\end{figure}

{\em Conclusions. - }
We show from covariant parton transport theory that in an expanding 
opaque quark-gluon plasma
there is a considerable probability that low-momentum
partons get accelerated to high transverse momenta in multiple scatterings.
This new mechanism of high-$p_T$ production, which is the opposite of jet 
quenching, has significant
implications for observables at RHIC energies and beyond.
For example, it gives a natural explanation to why elliptic flow at RHIC 
exceeds so called ``geometric'' bounds~\cite{shuryaklimit} and 
decreases relatively slowly at high transverse momenta.

This study considered six times higher opacities than that of a 
weakly-coupled perturbative parton 
plasma, which is 2-3 times smaller than the opacities 
estimated from RHIC data~\cite{v2} and likely even a smaller fraction of 
the opacities reachable at the future Large Hadron Collider (LHC).
At larger opacities, accelerated partons would have even larger importance
and would influence yields out to higher $p_T > 7-8$ GeV.
In addition, this work investigated only $2\to 2$ elastic and inelastic
transport. A more complete study will also have to consider in the future the 
effect of radiative 
$2\leftrightarrow 3$ processes.

{\em Acknowledgments. -}
Discussions with M.~Gyulassy are acknowledged.
This work was supported by DOE grant DE-FG02-01ER41190.

\end{document}